%% file: Main.tex
\documentclass[sigconf]{acmart}
\pdfoutput=1
\AtBeginDocument{%
  }

\copyrightyear{2026}
\acmYear{2026}
\setcopyright{cc}
\setcctype{by}
\acmConference[ICSE '26]{2026 IEEE/ACM 48th International Conference on Software Engineering}{April 12--18, 2026}{Rio de Janeiro, Brazil}
\acmBooktitle{2026 IEEE/ACM 48th International Conference on Software Engineering (ICSE '26), April 12--18, 2026, Rio de Janeiro, Brazil}
\acmPrice{}
\acmDOI{10.1145/3744916.3773232}
\acmISBN{979-8-4007-2025-3/2026/04}

\usepackage{listings}
\usepackage{amsmath}
\usepackage{amsfonts}
\usepackage{multirow}
\usepackage{graphicx}
\usepackage{subcaption} 

\usepackage{colortbl}
\usepackage{booktabs}
\usepackage{array}

\usepackage{tabularx}
\usepackage[linesnumbered,ruled,vlined]{algorithm2e}
\usepackage{color}

\newcommand{\tool}{\textsc{SAFE}}
\newcommand{\tooltwo}{\textsc{LCTGen}}
\newcommand{\toolthree}{\textsc{AC3R}}
\newcommand{\llmtool}{GPT-4o}

\begin{document}

\title{SAFE: Harnessing LLM for Scenario-Driven ADS Testing from Multimodal Crash Data}


\author{Siwei Luo}
\affiliation{%
  \institution{Macquarie University}
  \city{Sydney}
  \state{NSW}
  \country{Australia}}
\email{siwei.luo@hdr.mq.edu.au}

\author{Yang Zhang}
\affiliation{%
  \institution{University of North Texas}
  \city{Denton}
  \state{Texas}
  \country{USA}
}
\email{yang.zhang@unt.edu}

\author{Yao Deng}
\affiliation{%
  \institution{Macquarie University}
  \city{Sydney}
  \state{NSW}
  \country{Australia}
}
\email{yao.deng@hdr.mq.edu.au}

\author{Linfeng Liang}
\affiliation{%
  \institution{Macquarie University}
  \city{Sydney}
  \state{NSW}
  \country{Australia}
}
\email{linfeng.liang@hdr.mq.edu.au}

\author{Xi Zheng}
\authornote{Corresponding author}
\affiliation{%
 \institution{Macquarie University}
 \city{Sydney}
 \state{NSW}
 \country{Australia}
 }
 \email{james.zheng@mq.edu.au}

\renewcommand{\shortauthors}{Siwei et al.}

\begin{abstract}
Ensuring the safety of Autonomous Driving Systems (ADS) requires realistic and reproducible test scenarios, yet extracting such scenarios from multimodal crash reports remains a major challenge. Large Language Models (LLMs) often hallucinate and lose map structure, resulting in unrealistic road layouts and vehicle behaviors. To address this, we introduce \textbf{\tool}, a novel \textbf{S}cenario-based \textbf{A}DS testing \textbf{F}ramework via multimodal \textbf{E}xtraction, which leverages Retrieval-Augmented Generation (RAG), knowledge-grounded prompting, Chain-of-Thought (CoT) reasoning, and self-validation to improve scenario reconstruction from multimodal crash data. 

\tool{} achieves 93.8\% accuracy in extracting road network details, 80.0\% for actor information, and 100\% for environmental context. In human studies, \tool{} outperforms \tooltwo{} and \toolthree{} in reconstructing consistent road networks and vehicle behaviors. Under identical ADS and simulator settings, \tool{} detects 39 and 71 more safety violations than \tooltwo{} and \toolthree{}, respectively, and reproduces 12 more real-world crash cases than \tooltwo{}. On 19 cases supported by \toolthree{}, \tool{} reproduces one additional crash case with statistically significant gains across five runs. 
% It generates scenarios within 25 seconds and triggers violations after just 1 case (IDM), 3 cases (PPO), and 1 case (BeamNG). Unlike \toolthree{}, \tool{} is ontology-free and generalizes to a broader range of crash scenarios. 
It generates scenarios within 25 seconds and triggers violations after just 1 case (IDM) and 3 cases (PPO) in MetaDrive, as well as 1 case (Auto) in BeamNG.

% \textbf{Code:} \url{https://github.com/Siwei-Luo-MQ/SAFE-ADS-Testing}
\end{abstract}





\setcopyright{none} % to remove the copyright notice
\settopmatter{printacmref=false} % to remove the ACM Reference Format
\renewcommand\footnotetextcopyrightpermission[1]{}
\maketitle

\input{Introduction}

\input{Related_work}

\input{Methodology}

\input{Experiments}

\input{Results}

\input{Discussion}

\input{Conclusion}

\begin{acks}
This work is supported by the Australian Research Council grants FT240100269, LP190100676.
\end{acks}

\newpage
% \clearpage

\bibliographystyle{ACM-Reference-Format}
\bibliography{Main}

\end{document}

%% file: Introduction.tex
\section{Introduction}
Autonomous Driving Systems (ADS) are increasingly central to urban mobility, yet safety remains a critical concern due to recurring failures~\cite{ADAS_NHTSA_report}. Scenario-based simulation has become a scalable and cost-effective method for ADS testing~\cite{10.1145/3540250.3549111}, often relying on expert knowledge or pre-recorded data~\cite{nalic2020scenario}. Tools like Law-Breaker~\cite{sun2022lawbreaker}, RMT~\cite{deng2021rmt}, and TARGET~\cite{deng2023targetautomatedscenariogeneration} use traffic rules to generate scenarios, but lack detail in road geometry and vehicle behavior~\cite{TexasDMV2022}. Log-based approaches~\cite{LogSim} offer greater realism but face challenges in extracting critical corner cases from large-scale data.

To tackle these challenges, a distinct approach is proposed to reconstruct scenarios using real-world crash data. This method, initially introduced by ~\citet{gambi2019generating} with \toolthree{}, utilizes crash summaries to capture scenarios that challenge even experienced human drivers, thus providing highly pertinent cases for testing ADS. Crash data from the National Highway Traffic Safety Administration’s (NHTSA) CIREN dataset~\cite{CIREN} provides reliable, standardized accounts of crash events, offering insights into the underlying causes of critical incidents. Building on this, ADEPT~\cite{wang2022adept}, \tooltwo{}~\cite{tan2023language} and SoVAR~\cite{guo2024sovar} use crash summaries to reconstruct scenarios.

However, existing methods face key limitations: (1) relying solely on textual crash summaries omits critical visual context from crash sketches, such as vehicle positions and surroundings; (2) large language models (LLMs) used for information extraction suffer from hallucinations that are poorly managed; and (3) prompts are not tailored to specific crash cases, leading to irrelevant outputs that degrade extraction quality. Recent works leverage LLMs' multimodal capabilities, which unify tasks once handled by separate models like CLIP and BERT. \tooltwo{}~\cite{tan2023language} uses GPT-4~\cite{openai2024gpt4technicalreport} for more efficient and accurate extraction compared to traditional NLP in \toolthree{}~\cite{gambi2019generating}. LEAD~\cite{tian2024llm} and DeepCrashTest~\cite{bashetty2020deepcrashtest} further explore LLMs for video-based scenario configuration. Despite their strengths, LLMs remain vulnerable to hallucinations, producing plausible yet incorrect outputs.

Recent research highlights that hallucinations remain prevalent in LLMs, even with in-context learning~\cite{huang2023survey}, raising concerns about their reliability. To address limitations in extracting critical scenarios from crash reports—including map information loss, insufficient domain grounding, and hallucination—we propose \textbf{{\tool}}, a novel \textbf{S}cenario-based \textbf{A}DS testing \textbf{F}ramework via multimodal \textbf{E}xtraction. \tool{} introduces four key innovations: (1) a multimodal LLM pipeline that extracts information from both crash sketches and summaries with cross-validation; (2) a retrieval-augmented prompt generator that leverages pre-written templates and a domain-specific language (DSL) derived from real crash cases to provide contextual grounding; (3) a compact DSL, inspired by TARGET~\cite{deng2023targetautomatedscenariogeneration}, for accurately representing road networks, actors, and environments; and (4) a self-validator, inspired by SelfCheckGPT~\cite{manakul2023selfcheckgpt}, that combines in-context learning with Chain-of-Thought (CoT) prompting to reduce hallucination.
Extensive experiments demonstrate that \tool{} improves scenario extraction accuracy, test case alignment, and ADS bug detection, and reproduces more real-world crash cases compared to SOTA baselines. Our key contributions are:

\begin{itemize}
\item \textbf{ADS Test Case Generation from Multimodal Data:} Unlike prior work that relies solely on crash report summaries, \tool{} is the first to utilize both textual and visual crash data (summaries and sketches) with LLMs to improve scenario consistency. We also design a new DSL that encodes road geometry and traffic actor behaviors for precise scenario representation.

\item \textbf{Prompt Generation via Knowledge Base and Structured Prompting:} A prompt generator retrieves the most relevant prompt template and DSL oracle from a knowledge base based on the crash report, then applies CoT and few-shot prompting to construct robust, context-aware prompts. These prompts work with the Self-Validator to mitigate LLM hallucination and enhance extraction fidelity.

\item \textbf{Qualitative and Quantitative Analysis of Scenario Generation and Crash Reproduction:} Using 50 crash reports, \tool{} generates 521 scenarios across three ADSs and two simulators, identifying 159 critical cases on average. It detects 39 more safety violations than \tooltwo{} (IDM) and 71 more than \toolthree{} (Auto). User studies confirm better alignment with ground truth scenarios in terms of road layout and vehicle behavior. In crash reproduction, \tool{} reconstructs 12 more cases than \tooltwo{} out of 50, and 17 vs. 16 on a 19-case subset compared to \toolthree{}, both with statistically significant U-test results~\cite{WikipediaMann–WhitneyUtest}. Unlike ontology-constrained methods, \tool{} generalizes better across diverse crash types and real-world datasets. Code and experimental results are available in our \href{https://github.com/Siwei-Luo-MQ/SAFE-ADS-Testing}{GitHub repository}.
\end{itemize}

The remainder of the paper is organized as follows: Section ~\ref{sec:re_work} summarizes the related work. Section ~\ref{sec:methodology} details our proposed framework - {\tool}, while Section ~\ref{sec:experiments} describes the experimental setup. Section ~\ref{sec:results} presents an analysis of our findings. Section ~\ref{sec:discuss} discusses limitations, potential factors influencing our results and the future work. Finally, Section ~\ref{sec:conclusion} summarizes our work.

%% file: Related_work.tex
\section{Related Work}
\label{sec:re_work}
\subsection{Scenario-based ADS Testing}
Scenario-based testing is widely adopted for uncovering ADS safety issues, where domain knowledge plays a crucial role in efficient scenario construction and corner case detection~\cite{nalic2020scenario}. Tools like LawBreaker~\cite{sun2022lawbreaker}, RMT~\cite{deng2021rmt}, and TARGET~\cite{deng2023targetautomatedscenariogeneration} use traffic rules to define critical scenarios, but often lack realism due to missing details on road networks and environments~\cite{TexasDMV2022}.

M-CPS~\cite{zhang2023building} extracts scenarios from CCTV using semantic segmentation and object tracking, though video quality and event annotation gaps limit realism. LEAD~\cite{tian2024llm} and DeepCrashTest~\cite{bashetty2020deepcrashtest} use dashcam footage, but their narrow field of view fails to capture all traffic participants. \toolthree{}~\cite{gambi2019generating} initiated crash-report-based testing, leveraging structured sketches and text to capture roads, actors, and environments~\cite{CIREN}. Subsequent works such as ADEPT~\cite{wang2022adept}, \tooltwo{}~\cite{tan2023language}, and SoVAR~\cite{guo2024sovar} improved extraction accuracy via LLMs.

However, most prior methods rely solely on text, neglecting crucial sketch information such as road topology and vehicle trajectories, which leads to inaccurate scenario reconstruction. In contrast, our approach incorporates multimodal data and refined prompt engineering to improve extraction accuracy. We further introduce custom map generation to overcome the limitations of default simulator maps, enabling precise reconstruction of diverse crash scenes.

\subsection{Approaches to Mitigate LLM Hallucination}
While LLMs have shown strong generalization abilities, hallucination remains a major challenge, particularly in domain-specific tasks~\cite{huang2023survey}. Mitigation strategies include prompt engineering, retrieval-augmented generation (RAG), and validation mechanisms. CoT prompting~\cite{10.5555/3600270.3602070} improves reasoning by decomposing complex tasks into stepwise logic. Validation methods assess output reliability, such as Interrogate LLM~\cite{yehuda2024interrogatellm}, which checks answer-question consistency, and SelfCheckGPT~\cite{manakul2023selfcheckgpt}, which compares multiple responses for coherence. RAG~\cite{lewis2021retrievalaugmentedgenerationknowledgeintensivenlp} enhances accuracy by incorporating retrieved domain knowledge into prompts, reducing hallucinations and improving factual consistency.

This work leverages RAG by prebuilding a knowledge base of crash analysis templates and DSL examples to guide the LLM with domain-specific context. A prompt generator retrieves and assembles CoT- and few-shot-based templates tailored to each crash case. A self-validator further enhances reliability by checking consistency between the extracted information, the query, and source data. The modular, plug-and-play design enables each component to generalize across tasks.

%% file: Methodology.tex
\section{Methodology}
\label{sec:methodology}
\subsection{Overview}
\begin{figure}[!t]
  \centering
  \includegraphics[width=\linewidth]{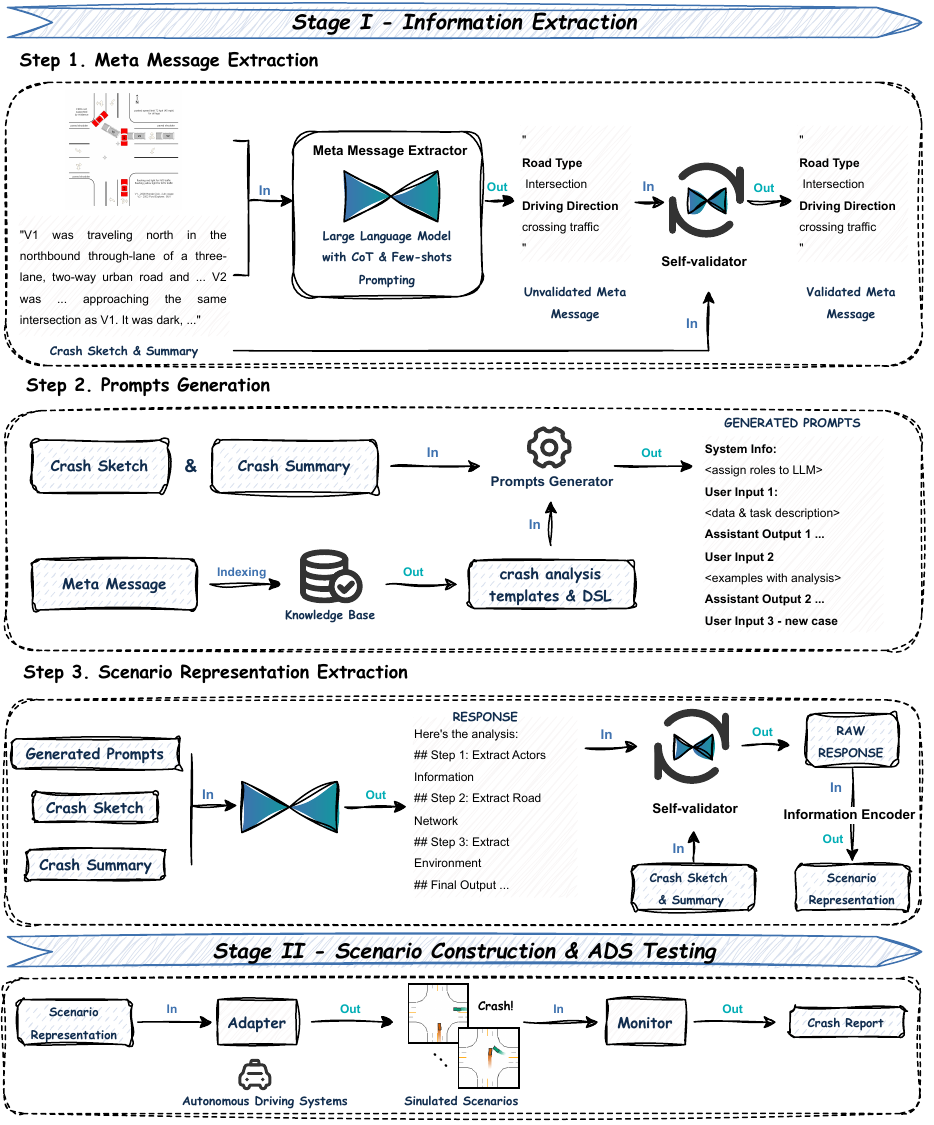}
  \Description{This figure illustrates the structure of SAFE, showing the pipeline from multimodal crash data extraction to scenario generation and ADS testing.}
  \caption{SAFE Structure}
  \label{fig:SAFE_structure}
\end{figure}

Figure ~\ref{fig:SAFE_structure} presents the whole structure of {\tool}, which includes two primary stages: the Information Extraction stage and the Scenario Construction \& ADS Testing stage. 
In Stage I, we develop a Meta Message Extractor, Self-Validator, and Prompts Generator to guide the LLM in extracting road, environment, and traffic actor details from multimodal crash reports and encoding them in the specified DSL.

Existing scenario-based ADS testing methods differ in representation, each with trade-offs. ADEPT~\cite{wang2022adept} uses Scenic~\cite{Fremont_2019}, but it often produces syntactically incorrect scripts, as shown in TARGET~\cite{deng2023targetautomatedscenariogeneration}, or syntactically correct yet semantically invalid scenarios, as further demonstrated in ScenicNL~\cite{elmaaroufi2024scenicnl}. \tooltwo{}~\cite{tan2023language} adopts high-dimensional Transformer-based encodings that improve structure but reduce interpretability and hinder validation. TARGET introduces a lightweight DSL, but oversimplifies road layouts (e.g., straight roads only, no lane info) and actor behaviors (e.g., relative but not absolute positioning), limiting realism.

To address this, we extend TARGET’s DSL with finer-grained attributes. For road networks, we add ‘Number of lanes’ and ‘Stem road direction’ (for T-intersections) to enable precise layout definition. For traffic actors, we refine positioning with directional labels (e.g., ‘W2E’ for a vehicle in the westernmost lane heading east). In complex layouts like merging roads, absolute labels (‘Main road’, ‘On-ramp’) replace ambiguous relative terms. 

This enhanced DSL (Listing~\ref{lst:DSL}) enables accurate, interpretable encoding of road structures and actor behaviors from crash reports, improving the syntactic and semantic fidelity of generated scenarios.

% https://drive.google.com/file/d/1t8BiZrq5cmxfJxHzCSLVIIIuGtFAreyj/view?usp=sharing

\begin{lstlisting}[caption=The Structure of Scenario DSL, label={lst:DSL}]
<Scenario>          ::=<Road network>;<Actors>;<Env>
<Road network>      ::=<Road type>;<No. lanes>;<Stem <@{\hspace*{9em}}@>direction>
<Road type>         ::=Straight | Curve | Intersection | T-<@{\hspace*{9em}}@>intersection | Merging
<No. lanes>         ::=Total number of lanes on the road
<Stem direction>    ::=North | South | East | West
<Actors>            ::=<Vehicle_1>;...;<Vehicle_n>
<Vehicle_n>         ::=<model>;<Initial_position>;<Actions<@{\hspace*{9em}}@>>;<Speed_limit>
<model>             ::=Sedan | SUV | Minivan | Pickup | Semi <@{\hspace*{9em}}@>Truck
<Initial_position>  ::=W2E | E2W | S2N | N2S | Main road | <@{\hspace*{9em}}@>On-ramp
<Actions>           ::=Move Forward | Turn Left | Turn Right
<Speed_limit>       ::=speed limit in mph
<Env>               ::=<Time>;<Weather>
<Time>              ::=Daytime | Nighttime
<Weather>           ::=Sunny | Cloudy | Overcast | Rainy | <@{\hspace*{9em}}@>Snowy | Foggy | Windy | Clear
\end{lstlisting}

In the second stage, we develop a scene generation adapter, which serves as a bridge between the scenario representation and the test scenarios executed in the simulator. It reconstructs the accident scene by invoking the simulator's map and environment module APIs based on the road network and environmental descriptions in the scenario representation. Additionally, it converts actor actions into ADS waypoints, thereby generating test scenarios. During testing, a dedicated scenario monitor tracks the system’s responses and generates a comprehensive test report detailing key evaluation metrics, such as the number of scenario builds, instances of collision scenarios, and overall ADS performance metrics at the end.

\subsection{Stage I: Information extraction}
Figure ~\ref{fig:SAFE_structure} 
Stage I consists of three main steps: Meta-Message Extraction, Prompts Generation, and Scenario Representation Extraction. The final output is a structured scenario representation, which serves as input for Stage II—Scenario Construction \& ADS Testing. An example scenario representation is shown in Listing~\ref{lst:scenario_representation}, derived from the crash report described in Figure~\ref{fig:SAFE_structure}, Stage I, Step 1.

\subsubsection{Step 1: meta message extraction}
Existing works, such as {\tooltwo}~\cite{tan2023language}, ADEPT~\cite{wang2022adept}, and SoVAR~\cite{guo2024sovar}, employ a one-time approach for extracting scenario information from accident reports using LLMs. They directly input crash summaries and prompts, which leads to several issues. Complex tasks and lengthy inputs degrade LLM performance, causing severe hallucinations~\cite{10.5555/3600270.3602070}. Additionally, the diverse nature of accident scenarios makes designing a single effective prompt challenging. 

To address these limitations, {\tool} introduces a two-stage scenario extraction process. In Stage I, the LLM extracts meta-information, such as road type and actor direction, from multimodal crash reports. In Stage II, specialized prompts and few-shot examples refine the input, improving extraction accuracy. This approach leverages multimodal LLMs, few-shot learning, CoT prompting, and self-validation to enhance robustness.

Table~\ref{tab:meta_msg_prompts} outlines the meta-message extraction prompts, which consist of five key components: System Info, User Input 1–3, and Assistant Output 1–3. These prompts are used to drive the LLM's reasoning and extraction process, forming the core of the Meta Message Extractor module. To ensure consistency and clarity, we structure these prompts using the OpenAI Chat Message Format~\cite{openai_chatformat}. \textit{System Info} sets global LLM instructions, guiding response style and behavior~\cite{Persona}, instructing the LLM to act as an expert road engineer. \textit{User Input 1} defines the task, analysis steps, and expected output, followed by \textit{Assistant Output 1}, linking task requirements to case analysis. \textit{User Input 2} employs CoT to decompose tasks into subtasks, validated using crash sketches and summaries. \textit{Assistant Output 2} reinforces objectives, while \textit{User Input 3} provides a new case, prompting \textit{Assistant Output 3} to generate scenario representations. To ensure self-consistency, an LLM-driven self-validator integrates previous responses with the original queries to verify the Meta Message Extractor’s outputs. If inconsistencies are found, the extraction process is re-executed.

\begin{lstlisting}[caption=A Scenario Representation Example - case 117021, label={lst:scenario_representation}]
<Scenario>:
    <Road network>:
        <Road type>: Intersection
        <No. lanes>: 3
        <Stem direction>: Not applicable
    <Actors>:
        <Vehicle_1>:
            <Model>: Sedan
            <Initial_position>: S2N
            <Actions>: Move forward
            <Speed_limit>: 45
        <Vehicle_2>:
            <Model>: SUV
            <Initial_position>: E2W
            <Actions>: Move forward
            <Speed_limit>: 45
    <Env>:
        <Time>: Nighttime
        <Weather>: Clear
\end{lstlisting}

\subsubsection{Step 2: prompts generation}
By analyzing the original data~\cite{CIREN}, we observed that crash cases that occur on the same type of road share many commonalities in scenario descriptions and definitions, whereas cases on different types of roads exhibit significant variations in how they are described and defined. A simple example is that for accidents occurring on straight roads, vehicle positions can be described using directional phrases such as "traveling east to west." However, for crashes on merging segments, it is crucial to first distinguish whether the vehicle is on the main road or the on-ramp. When using LLMs, including examples or definitions from other road types within prompts designed for straight roads can negatively impact the extraction accuracy. To address this challenge and provide more precise and concise prompts, {\tool} introduces the concept of a knowledge base and prompts generation, inspired by RAG~\cite{lewis2021retrievalaugmentedgenerationknowledgeintensivenlp}. RAG is a technique that combines information retrieval and text generation, leveraging external knowledge bases or documents to enhance the accuracy of LLM response and expand knowledge coverage, thus reducing hallucination issues. Following this approach, we construct an external knowledge base and retrieval system, tailored to the five road types covered in crash reports: Straight, Curve, Intersection, T-intersection, and Merging. For each road type, we develop crash analysis templates and index the knowledge base using the type of road and the direction of the traffic actor. Each crash analysis template consists of an actual crash case and a data analysis prompt designed using the CoT method. Once the meta message of a given crash case is determined, a prompts generator retrieves relevant example data and analysis templates from the knowledge base, constructing optimized prompts for information extraction. Table ~\ref{tab:crash_temp} presents a crash analysis template for intersection crashes, which is built upon a real-world crash case and details how to analyze traffic actors' behavior, road network, and environmental information. This template is stored in the knowledge base and can be indexed by road type. Table ~\ref{tab:sr_extract} provides a fully formed prompt that incorporates relevant information retrieved from the knowledge base, demonstrating how optimized prompts are generated for LLM-based crash analysis.

\begin{table}[!t]
\centering
\resizebox{\columnwidth}{!}{
\begin{tabular}{p{0.9\columnwidth}}
\toprule
\rowcolor{gray!20} System Info \\

Role: An experienced road engineering expert skilled in identifying road types.\\

\rowcolor{gray!20} User Input 1 \\

1. Dataset Description:

(1)Crash Sketch: Depict road networks, vehicle trajectories...

(2)Crash Summary: Describe the road networks, and vehicle actions...

2. Task definition:

(1)Extract the Road Network: ...

(2)Identify the driving direction of traffic actors: ...

3. Analysis steps:

\#\#\#Step 1. Extract the road network

\#\#\#\#Step 1.1 Read the crash summary to identify the content that describes the road segment

\#\#\#\#Step 1.2 Determine the road type based on the description.

\#\#\# Step 2 - Extract the number of cars involved in the crash:

\#\#\# Step 3 - Determine the initial driving direction of the cars:

4. Output Format:

\{'Road type': '<your answer>', 'Number of cars': <your answer>, 'Driving direction': '<your answer>'\}\\
\rowcolor{gray!20} Assistant Output 1 \\
Please go ahead and provide an example so I can better understand the task and assist you accordingly.\\
\rowcolor{gray!20} User Input 2 \\
Example <Sketch> \& <Summary>

Analysis Process:

xxx xxx (detailed analysis process for this example according to the given analysis steps)\\
\rowcolor{gray!20} Assistant Output 2\\
Got it! Please provide a new crash case and I'll follow the process to extract the required information.\\
\rowcolor{gray!20} User Input 3\\
New Case <Sketch> \& <Summary>\\
\bottomrule
\end{tabular}%
}
\caption{The Structure of Meta Message Extraction Prompts}
\label{tab:meta_msg_prompts}
\end{table}

\begin{table}[!t]
\centering
\resizebox{\columnwidth}{!}{
\begin{tabular}{p{0.9\columnwidth}}
\toprule
\rowcolor{gray!20} Input \\
Example Sketch \& Summary

DSL Template(Described in Listing~\ref{lst:DSL})\\
\rowcolor{gray!20} Analysis Process \\
\#\# Step 1: Extract Actors Information

1. \textbf{Vehicle Entity Recognition}

2. \textbf{Vehicle Model Recognition}
    
    - Vehicle\_1: ...
    
    - Vehicle\_2: ...

3. \textbf{Vehicle Movement Behavior Recognition}

\#\# Step 2: Extract Road Network

- The summary indicates a two-lane, two-way residential roadway.

- Number of Lanes: 2

\#\# Step 3: Extract Environment

- The summary states “It was daylight, the sky was cloudy”

- Time: Daytime

- Weather: Cloudy

\#\# Final Output

(formatted output based on the given DSL)\\
\bottomrule
\end{tabular}%
}
\caption{Crash Analysis Template for Intersection Crashes}
\label{tab:crash_temp}
 % \vspace{-2mm}
\end{table}

\begin{table}[!t]
\centering
\resizebox{\columnwidth}{!}{
\begin{tabular}{p{0.9\columnwidth}}
\toprule
\rowcolor{gray!20} System Info \\
Role: An experienced road engineering expert skilled in extracting scenario representations.\\
\rowcolor{gray!20} User Input 1 \\
Data \& Task: Extract scenario representations from crash reports according to the keys in DSL.

(Road-aware prompt retrieved from the knowledge base using the road type as the index.)
% (DSL indexed from \textbf{Knowledge Base} as shown in Listing~\ref{lst:DSL})

Output format: (provide formatted output)

Example: 

\textbf{Input}:

Crash summary - "A 2010 Mercedes-Benz C200 ...

Crash sketch - A sketch of the accident ...

\textbf{Output}: (formatted output)\\
\rowcolor{gray!20} Assistant Output 1 \\
Got it! Please provide me with an example so I can better understand this task!\\
\rowcolor{gray!20} User Input 2 \\
% (The Crash Analysis Template, crash sketch and summary, and DSL indexed from \textbf{Knowledge Base} as shown in Table ~\ref{tab:crash_temp})\\
(Road-aware prompt retrieved from the knowledge base using the road type as the index.)\\
\rowcolor{gray!20} Assistant Output 2 \\
Got it! Please provide a new crash case and I'll follow the process to extract the required information.\\
\rowcolor{gray!20} User Input 3 \\
Crash Sketch and summary form a new case\\
\bottomrule
\end{tabular}%
}
\caption{The Structure of Scenario Representation Extraction Prompts}
\label{tab:sr_extract}
 % \vspace{-10mm}
\end{table}

\subsubsection{Step 3: scenario representation extraction}
In the third step, {\tool} utilizes the LLM along with the specific prompts (shown in Table~\ref{tab:sr_extract}) generated in the second step for the current crash case to extract scenario information. To further reduce hallucinations and ensure output consistency, we apply the self-validator module. It re-feeds the extracted output, original query, and raw data into the LLM to verify alignment with the source. If validated, the output is passed to an information encoder, which uses regular expressions~\cite{WikipediaRegularExpression} to map data to DSL elements; otherwise, extraction is re-executed. This results in the generation of a complete scenario representation. Algorithm~\ref{alg:encoder} provides a detailed breakdown of the process, illustrating the entire workflow from Step 1 to Step 3.

\begin{algorithm}

\caption{Information Extraction}
\label{alg:encoder}

\KwIn{Crash Sketch \& Summary}
\KwOut{Scenario Representation (recorded in DSL)}

Meta message = LLM(Crash Sketch, Summary, Prompts ~\ref{tab:meta_msg_prompts})

Self validation = LLM(Crash Sketch, Summary, Meta message)

\If {Self validation != 'Pass'}{
    Meta message = LLM(Crash Sketch, Summary, Prompts ~\ref{tab:meta_msg_prompts})
}

Crash analysis template ~\ref{tab:crash_temp}, DSL = Knowledge Base Indexing(Meta message)

Prompts ~\ref{tab:sr_extract} = Prompts Generator(Crash Sketch, Summary, Crash analysis template ~\ref{tab:crash_temp}, DSL)

Raw response = LLM(Crash Sketch, Summary, Prompts ~\ref{tab:sr_extract})

Self validation = LLM(Crash Sketch, Summary, Raw response)

\If {Self validation != 'Pass'}{
    Raw response = LLM(Crash Sketch, Summary, Prompts ~\ref{tab:sr_extract})
}

Scenario representation = Information Encoder(Raw response) \tcp*{Regular Expression-based parser}

\Return{Scenario representation}
\end{algorithm}
% \vspace{-5mm}
% \vspace{-2mm}
\subsection{Stage II: Scenario Construction \& ADS Testing}

In Stage II, we generate test scenarios in a simulator using the extracted scenario representation and evaluate them against the chosen ADS. A monitoring system then produces a detailed test report, capturing scenario construction details and collision outcomes. To ensure the simulated road network aligns with the crash report, we employ a scene generation adapter that invokes the relevant simulator APIs to construct maps and integrate environmental and actor data into an executable script.

Simulators support two primary map creation methods: (1) \textit{Parameter-Based Construction}, which programmatically defines the road network using attributes such as road type, direction, lane count, and length, enabling structured and automated scenario generation; and (2) \textit{Build-From-Scratch}, where users manually design roads on a blank plane via a coordinate system, offering flexibility but requiring more effort and increasing inconsistency risks. By leveraging structured map construction and the scene generation adapter, our framework enhances the realism, accuracy, and reproducibility of ADS testing. In this paper, we selected MetaDrive~\cite{li2022metadrive} and BeamNG~\cite{beamng_tech} as our target simulators for scenario construction and testing. MetaDrive follows the first approach, while BeamNG follows the second approach in building maps.

\begin{figure}[!t]
  \centering
  \includegraphics[width=\linewidth]{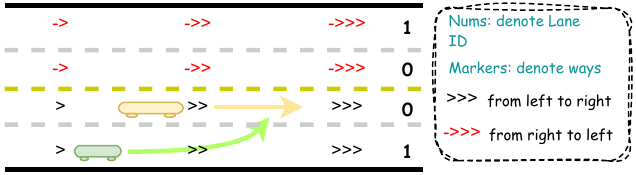}
  \Description{This figure illustrates how vehicle positions and related state information are aligned with the road map annotations in MetaDrive, showing the correspondence between the road network and the simulated vehicle trajectories.}
  \caption{Straight Road Network in MetaDrive}
  \label{fig:metadrive}
\end{figure}

Figure ~\ref{fig:metadrive} illustrates how a straight-road scene is constructed in the first-type simulator, using MetaDrive as an example. In MetaDrive, a straight-line lane for left-to-right direction consists of three key points, indicated as $>$, $>>$, and $>>>$, representing the beginning, middle, and end of the lane, respectively. Each lane in MetaDrive is assigned a unique ID, where `0` denotes the lane closest to the centerline, and higher numbers correspond to lanes farther away. To differentiate directions, while straight lanes running from left to right are marked with $>$, $>>$, and $>>>$,  those running from right to left use $<$, $<<$, and $<<<$ as lane markers. Based on this road structure, our developed Scene Generation Adapter maps vehicle actions extracted from the first stage to these lane IDs and lane markers, considering the number of lanes specified in the scenario representation. For constructing scenarios in the second type of simulator, the Adapter leverages road network descriptions from the scenario representation with predefined lane factors such as the width and length of a lane to generate a configuration file that defines the simulation environment. Once the road network is constructed in BeamNG, the Adapter maps the vehicle positions and behaviors described in the DSL to specific coordinate points. For example, a vehicle traveling west to east in a straight direction is mapped to three coordinate points along the west-to-east road, ensuring that the vehicle's behavior aligns with the DSL description. Unlike {\tooltwo}~\cite{tan2023language}, which randomly assigns an actor as the ego vehicle, our approach iterates through all actors in the scene, assigning each one as the ego car in turn. This method allows for the exploration of a broader range of critical scenarios. Each designated ego vehicle is then integrated with the ADS algorithm for testing. During the simulation, a monitor constantly tracks the test, records all scenarios, and creates a report detailing any detected collisions.

%% file: Experiments.tex
\section{Experiments}
\label{sec:experiments}
% Section - 4.1
\subsection{Research Questions}
We propose four research questions (RQs) along with related experiments to assess the effectiveness of {\tool}. The RQs are as follows:

\begin{itemize}
\item RQ1: How accurate is {\tool} in extracting scenario representations during the Information Extraction stage?

\item RQ2: How accurately do the critical scenarios constructed by {\tool} reflect the original crash report?

\item RQ3: How effectively can {\tool} construct scenarios from existing data, uncover ADS bugs and reproduce crashes?

\item RQ4: How effective are the proposed prompt engineering and validation methods?
\end{itemize}
% Section - 4.2
\input{Experiment_settings}
% Section - 4.3
\input{Evaluation_metrics}

%% file: Experiment_settings.tex
\subsection{Experiment Settings}
\subsubsection{Generic Settings}
To evaluate \tool{}, we extended the dataset from \tooltwo{}~\cite{tan2023language} by adding 12 crash reports to the original 38 from the NHTSA CIREN database, resulting in 50 cases. This full dataset was used to compare \tool{} with \tooltwo{} under identical conditions. \toolthree{} supports only rear-end and sideswipe crashes and is sensitive to input format due to its reliance on traditional NLP; thus, it could process only 19 of the 50 cases. Accordingly, comparisons with \toolthree{} were conducted on this 19-case subset. Each report includes a sketch of the crash scene and a textual summary of driver behavior, vehicle state, and environmental context. We use \llmtool{}~\cite{GPT-4o} for its SOTA multimodal capabilities and also replace GPT-4 with GPT-4o in \tooltwo{} to ensure a fair comparison.
%To evaluate {\tool}'s performance, we expanded the dataset from {\tooltwo}~\cite{tan2023language} by adding 12 crash reports to the original 38 from the NHTSA CIREN database, creating a dataset of 50 crash cases. This full set of 50 cases was used to evaluate {\tool} and to conduct a direct comparison with {\tooltwo} under identical conditions. We also contacted the authors of {\toolthree} and learned that the tool only supports rear-end and sideswipe crashes and is sensitive to input format due to its reliance on traditional NLP techniques. As a result, {\toolthree} could only successfully process 19 out of the 50 crash reports in our dataset. Therefore, the comparative evaluation between {\tool} and {\toolthree} was conducted on this 19-case subset. Each report contains a sketch illustrating the crash scene, including the map layout and vehicle trajectories, along with a summary detailing driver behavior, vehicle status, and environmental conditions. For our framework’s LLMs, we choose {\llmtool}~\cite{GPT-4o} because of its SOTA multimodal capabilities. To ensure a fair comparison and eliminate performance differences arising from the choice of LLM, we also replace the GPT-4 model used in {\tooltwo} with GPT-4o in our experiments.
 To test {\tool}’s scenario scalability and bug-detection capacity in ADS, we test {\tool} on two simulators: MetaDrive~\cite{li2022metadrive} and BeamNG~\cite{beamng_tech}. MetaDrive, a lightweight ADS testing simulator developed by UCLA, allows for customizable road scenarios and supports multiple ADS types—IDM (maintains safe distances with reinforcement learning algorithms) and PPO (end-to-end neural network model)~\cite{MetaDriveADS}. In contrast, BeamNG, a realistic driving simulator on the Torque3D engine, offers detailed vehicle models and customizable environments. On BeamNG, we test the auto-driving model, a widely used ADS among more than 250,000 Steam users, which supports advanced autonomous driving functionalities such as obstacle avoidance and lane switching.

\subsubsection{Settings for RQ1}
\label{sec:rq1}
To evaluate the accuracy of {\tool} in scenario representation extraction, we implement a validation process. We first randomly sampled 25\% cases from the total dataset as the test set. Two experimenters with expertise in ADS testing were independently hired to construct scenario representations for the test sample set based on the definitions of various attribute fields in the DSL. The definition of the DSL and its fields can be found in the Knowledge base. The experimenters then cross-checked each other's scenario representations, discussed and resolved any ambiguities, and ultimately produced a mutually agreed-upon scenario representation. This final version served as the human oracle for evaluating the results of the scenario representation extraction step in {\tool}.

Subsequently, we developed a validator program to compare the scenario representations generated by {\tool} with the human oracle on a field-by-field basis. We then calculated Road Network Accuracy, Actor Accuracy, Env Accuracy, and Overall Accuracy. Let \( N \) be the sample size, and define:

\begin{align}
\text{Road Network Accuracy} &= \frac{N_{\text{road}}}{N},  \\
\text{Actor Accuracy} &= \frac{N_{\text{actor}}}{N}, \\
\text{Env Accuracy} &= \frac{N_{\text{env}}}{N}, \\
\text{Overall Accuracy} &= \frac{N_{\text{all}}}{N}
\end{align}

where \( N_{\text{road}} \), \( N_{\text{actor}} \), \( N_{\text{env}} \), and \( N_{\text{all}} \) represent the number of samples where all fields in the respective categories are correctly extracted. These metrics were used to assess whether SAFE exhibited hallucinations and to evaluate the accuracy of the extracted information.

\subsubsection{Settings for RQ2}
To investigate whether the critical scenarios generated by {\tool} align with the scenarios described in the original crash reports and to analyze the differences between {\tool} and existing methods, we conducted a user study. A survey is designed in which participants are presented with the testing scenarios generated by {\tool} and existing methods on the test sample set, and the original test set data. Thirty participants, recruited globally through Prolific~\cite{Prolific} to ensure broad representativeness, are asked to evaluate the extent to which the testing scenarios generated by different approaches align with the scenarios described in the original crash reports. The evaluation is based on two aspects: road network and actor behaviors alignment, using a five-level rating scale: Totally match, Mostly match, Neutral, Mostly not match, and Totally not match. 
The original survey forms are available at: \href{https://forms.gle/vDggkGth9y3yw6498}{{\tool} V.S. {\tooltwo}}, and \href{https://forms.gle/ksSSAniXKcR89mbT9}{{\tool} V.S. {\toolthree}}. User study demographics are provided in our \href{https://github.com/Siwei-Luo-MQ/SAFE-ADS-Testing}{GitHub repository}. All tools were anonymized, and their order was randomized across questions to minimize bias. Regarding the selection of baselines, ADEPT~\cite{wang2022adept} uses DaVinci-003 as its LLM, a model that has long been surpassed by newer models and is no longer supported by the OpenAI API. SoVAR~\cite{guo2024sovar} employs the LGSVL~\cite{rong2020lgsvl} simulator to construct test scenarios; however, this simulator was discontinued in Jan 2022, and its servers have been shut down. Therefore, we selected {\tooltwo}~\cite{tan2023language}, the current SOTA method, and {\toolthree}~\cite{gambi2019generating} as two baselines for evaluating scenario reconstruction consistency.

\subsubsection{Settings for RQ3}
To evaluate \tool{} in scenario construction and ADS bug detection, we applied \tool{}, \tooltwo{}~\cite{tan2023language}, and \toolthree{}~\cite{gambi2019generating} to all 50 crash reports to extract scenario representations. Using consistent simulators (MetaDrive, BeamNG) and ADS models (IDM, Auto), we compared the number of generated scenarios, detected bugs, execution time, and Top-K bug coverage. We also included PPO in MetaDrive to further assess \tool{}’s capability. For crash reproduction, we fixed the vehicle spawn seed for consistent scenario generation. \tool{} and \tooltwo{} were evaluated on all 50 reports using MetaDrive and IDM. Since \toolthree{} supports only a subset of crashes, we compared \tool{} and \toolthree{} on 19 reports using BeamNG and the Auto model, measuring successful reproduction rates.
%To evaluate {\tool}’s performance in scenario construction and ADS bug detection compared to competing techniques, we first applied {\tool}, {\tooltwo}~\cite{tan2023language}, and {\toolthree}~\cite{gambi2019generating} to the entire dataset of 50 crash reports to extract scenario representations. For a fair comparison, we used the same simulators (MetaDrive and BeamNG) and ADS models (IDM and Auto) for scenario construction and testing, and reported metrics including the number of generated scenarios, detected bugs, execution time, and Top-K bug coverage. Additionally, we tested the PPO ADS model using MetaDrive to further assess {\tool}’s capability.

%To evaluate {\tool}’s ability to reproduce real-world crashes, we fixed the random seed for vehicle spawn point selection to ensure consistent scenario generation. We then applied both {\tool} and {\tooltwo} to all 50 crash reports using the MetaDrive simulator and IDM model, and measured how many crashes could be successfully reproduced. For {\toolthree}, which only supports a subset of crash types, we applied both {\tool} and {\toolthree} to the 19 supported crash reports using the BeamNG simulator and Auto ADS model, and similarly measured reproduction success rates.
\subsubsection{Settings for RQ4}
To evaluate the effectiveness of our proposed prompt engineering, knowledge base and validation methods, we perform ablation studies. First, we remove the prompts generation from Stage I and measure the accuracy metrics mentioned in section ~\ref{sec:rq1} on the sampled test data. Next, we remove the self-validation process and measure the scenario representation accuracy under the same metrics again.

%% file: Evaluation_metrics.tex
\subsection{Evaluation Metrics}
In RQ1, we evaluate four key metrics: Road Network Accuracy, Actor Accuracy, Environment Accuracy, and Overall Accuracy.
%In RQ1, we consider the following statistics:
%\begin{itemize}
%\item Road Network Accuracy
%\item Actor Accuracy
%\item Env Accuracy
%\item Overall Accuracy
%\end{itemize}
In RQ2, we present the results of the ``Scenario Reconstruction User Study," showing rating distributions for each tool across different questions. We also provide a comparative analysis of \tool{} against \tooltwo{} and \toolthree{} on scenario generation performance. 
%In RQ2, we present the performance of {\tool}, {\tooltwo} and {\toolthree}, including the distribution of ratings across different cases for different questions from the ``Scenario Reconstruction User Study''. Furthermore, we provide a comparative analysis of {\tool} and two baselines on scenario generation performance.
In RQ3, we evaluate and compare \tool{}, \tooltwo{}, and \toolthree{} using the following statistics: the number of scenarios generated, number of violations detected, scenario generation time, number of scenarios required to find the Top-$k$ violations (for $k=1,2,3$), the average violation detection ratio, and the number of successfully reproduced crashes.
%In RQ3, we consider the following statistics:
%\begin{itemize}
%\item Number of scenarios generated by {\tool}, {\tooltwo} and {\toolthree}
%\item Number of violations detected by {\tool}, {\tooltwo} and {\toolthree}
%\item Scenario generation time of {\tool}, {\tooltwo} and {\toolthree}
%\item Number of scenarios used to find the Top-k violation of {\tool}, {\tooltwo} and {\toolthree} (where $k$ is set to 1, 2, and 3 in this paper).
%\item Average ratio of finding violations of {\tool}, {\tooltwo} and {\toolthree}.
%\item Number of crashes reproduced by {\tool}, {\tooltwo} and {\toolthree}
%\end{itemize}
In RQ4, we consider the same metrics as RQ1.

%% file: Results.tex
\section{Results}
\label{sec:results}
\subsection{RQ1: Accuracy of Scenario Extraction}
Table~\ref{tab:safe_performance} reports \tool{}’s average scenario extraction performance across five runs. \tool{} consistently achieves 93.8\% Road Network Accuracy, 100\% Environment Accuracy, and 80.0\% accuracy for both Actor and Overall metrics, confirming its effectiveness in extracting structured scenarios from multimodal crash reports.

Analysis of actor extraction failures reveals that most errors arise from challenges in identifying vehicle types and initial positions. For example, when a vehicle is described as ``a 2004 Jeep Grand Cherokee'', the LLM may misclassify it as a sedan rather than an SUV, reflecting ambiguity in vehicle categorization. Positioning errors were also evident, such as in crash report 120523 (Figure~\ref{fig:sc_safe_lctgen}, top-right), where the LLM incorrectly described Vehicle~1 as traveling west to east and entering the wrong way. In reality, the sketch shows it moving east to west. We attribute such discrepancies to the low resolution of some sketches, which reduces visual clarity and may mislead the LLM during interpretation.

These issues may be addressed by refining the DSL schema and incorporating annotated driving data into the RAG process. We discuss these limitations and mitigation strategies further in Section~\ref{sec:discuss}. Overall, \tool{} demonstrates strong and reliable performance in scenario representation and reduces hallucinations in LLM-based extraction.

\begin{table}[!ht]
    \centering
    \renewcommand{\arraystretch}{1.2}
    \begin{tabular}{lcc}
        \hline
        \textbf{Metric} & \textbf{Mean} & \textbf{Std.}\\
        % \hline
        Road Network Accuracy & 93.8\% & 3.4 \\
        % \hline
        Actors Accuracy & 80.0\% & 10.3 \\
        % \hline
        Env Accuracy & 100\% & 0.0 \\
        % \hline
        Overall Accuracy & 80.0\% & 10.3 \\
        \hline
    \end{tabular}
    \caption{Performance of SAFE in Scenario Representation Extraction}
    \label{tab:safe_performance}
    % \vspace{-8mm}
\end{table}

\subsection{RQ2: Scenario Alignment Evaluation}
To assess scenario consistency and compare \tool{} with \tooltwo{} and \toolthree{}, we conducted a user study. For \tool{} vs. \tooltwo{}, participants evaluated reconstructed scenarios from 13 randomly selected crash reports; for \tool{} vs. \toolthree{}, 19 reports supported by \toolthree{} were used. In both cases, participants rated alignment with the original crashes based on road network and actor vehicle behavior.

\begin{figure}[t]
  \centering
  \begin{subfigure}{\linewidth}
      \centering
      \includegraphics[width=0.9\linewidth]{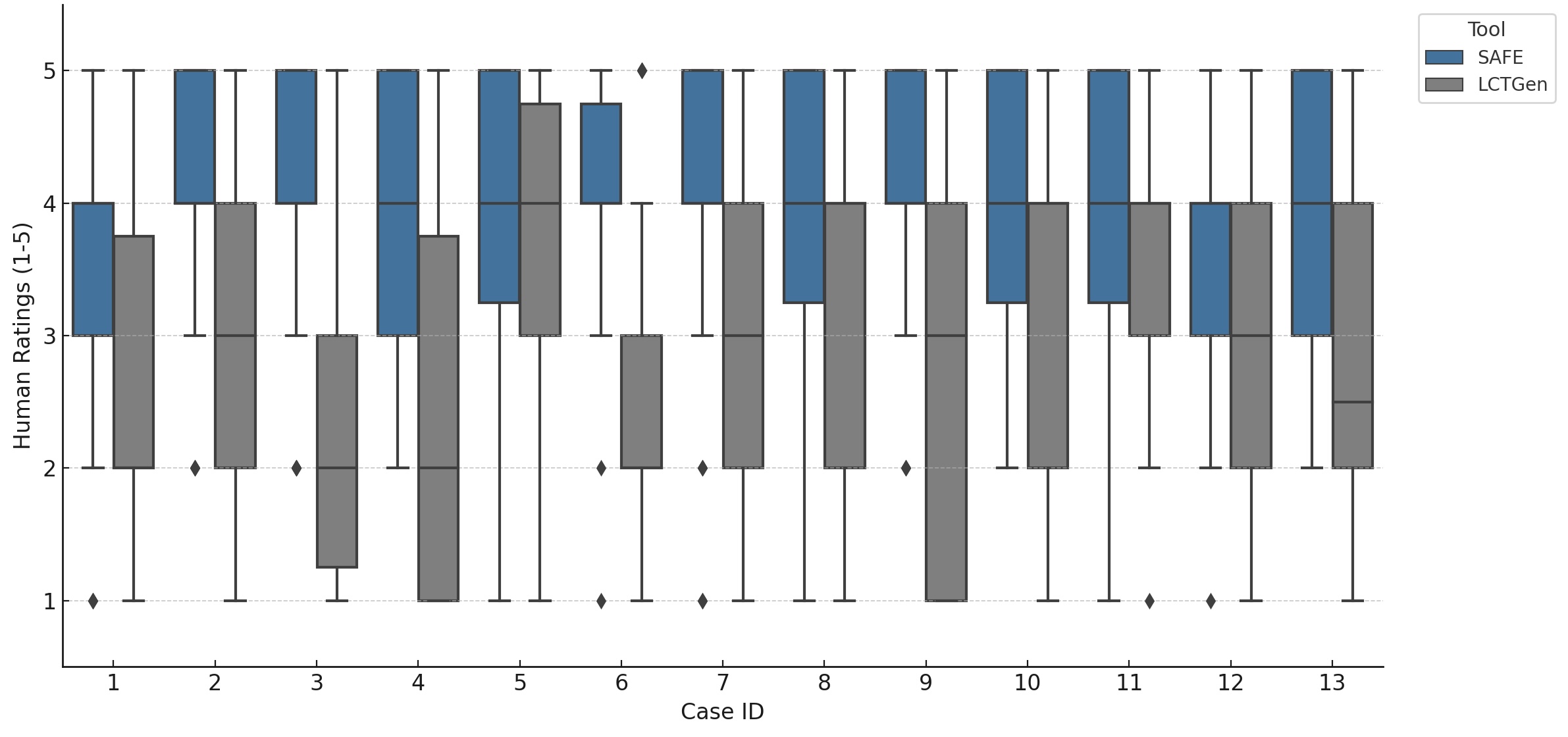}
      \Description{This subfigure presents the human evaluation of road network reconstruction consistency between SAFE and LCTGen, comparing how accurately each method reconstructs road structure.}
      \caption{Road Network Alignment Comparison}
      \label{fig:subfig_a_1}
  \end{subfigure}
    
  \begin{subfigure}{\linewidth}
      \centering
      \includegraphics[width=0.9\linewidth]{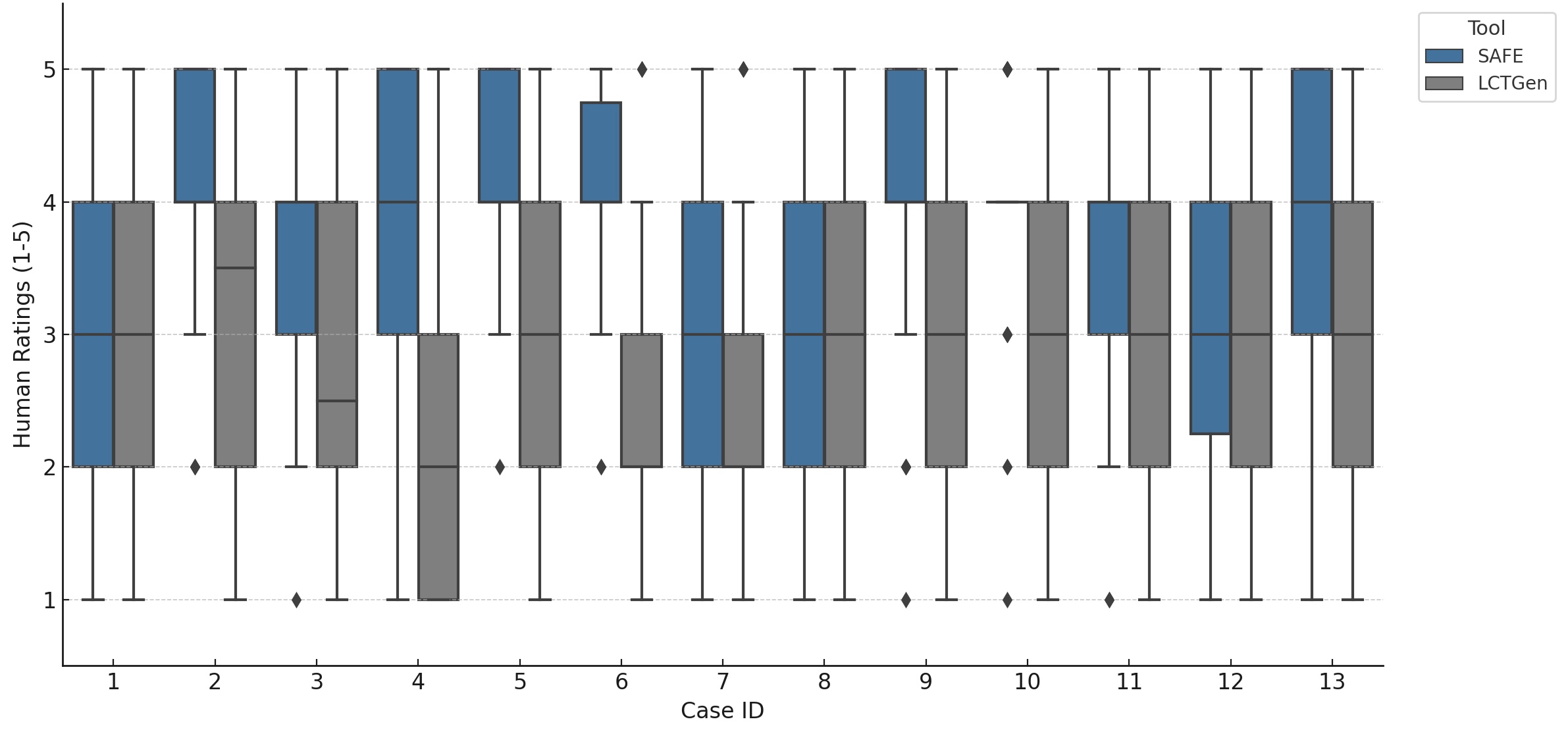}
      \Description{This subfigure presents the human evaluation of actor behaviour reconstruction consistency between SAFE and LCTGen, assessing how well each method reproduces vehicle movements and interaction patterns.}
      \caption{Actor Behaviours Alignment Comparison}
      \label{fig:subfig_a_2}
  \end{subfigure}
  \caption{{\tool} vs. {\tooltwo} Human Study Results}
  \label{fig:safe_lctgen_human_study}
\end{figure}

Figure~\ref{fig:safe_lctgen_human_study} shows the user study results comparing \tool{} and \tooltwo{}. \tool{} significantly outperforms \tooltwo{} in both Road Network Alignment and Actor Behavior Alignment based on mean scores. Mann–Whitney U tests on ratings from 30 participants across 13 cases confirm this gap: for road alignment, $U=106{,}546.5$ with $p=6.85 \times 10^{-24}$; for actor behavior, $U=99{,}612.0$ with $p=1.46 \times 10^{-14}$. These results strongly support \tool{}’s superiority in scenario consistency.

\begin{figure}[!ht]
  \centering
  \includegraphics[width=\linewidth]{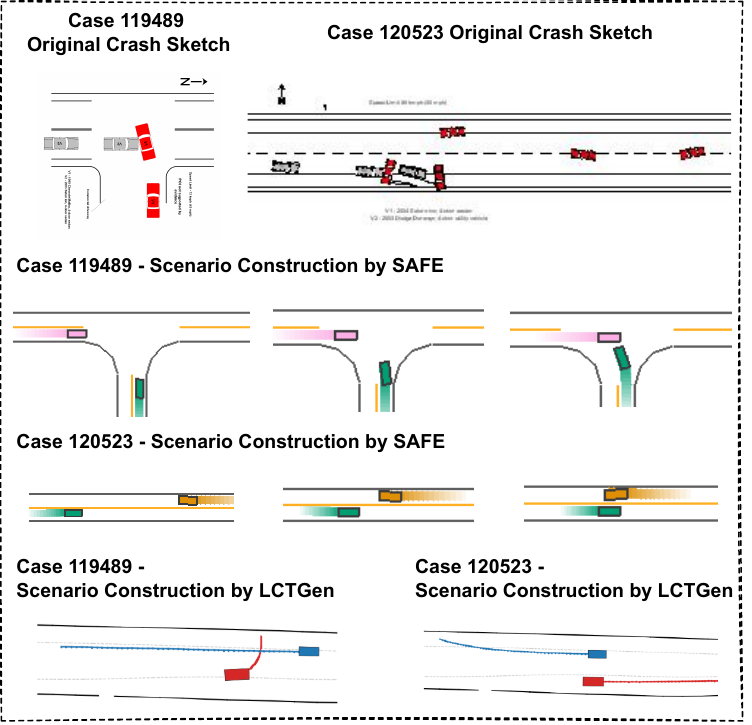}
  \Description{This figure shows two examples regarding SAFE V.S. LCTGen on Scenario Construction}
  \caption{{\tool} vs. {\tooltwo} on Scenario Construction}
  \label{fig:sc_safe_lctgen}
\end{figure}

Figure~\ref{fig:sc_safe_lctgen} compares scenario construction results from \tool{} and \tooltwo{} for Crash Case 119489 (Case 9) and 120523 (Case 11) from the user study. In Case 119489, \tool{} correctly reconstructs a bidirectional two-lane T-intersection and accurately restores actor behaviors, including the ego vehicle’s failed yielding attempt. In contrast, \tooltwo{} misclassifies the road as a straight segment and produces an incorrect actor behavior sequence. In Case 120523, \tool{} misinterprets the vehicle direction due to low-resolution sketch input, leading to a behavior extraction error, while \tooltwo{} places vehicles in the wrong lane but fails to create a critical interaction. Notably, \tool{} still reconstructs the correct road type, whereas \tooltwo{} generates an incorrect three-way configuration. Overall, the user study and case analysis highlight that \tooltwo{} struggles with road and actor accuracy, while \tool{} benefits from its detailed DSL, multimodal inputs, and LLM-driven extraction to produce more realistic scenarios and better support bug detection.

\begin{figure}[t]
  \centering
  \begin{subfigure}{\linewidth}
      \centering
      \includegraphics[width=0.9\linewidth]{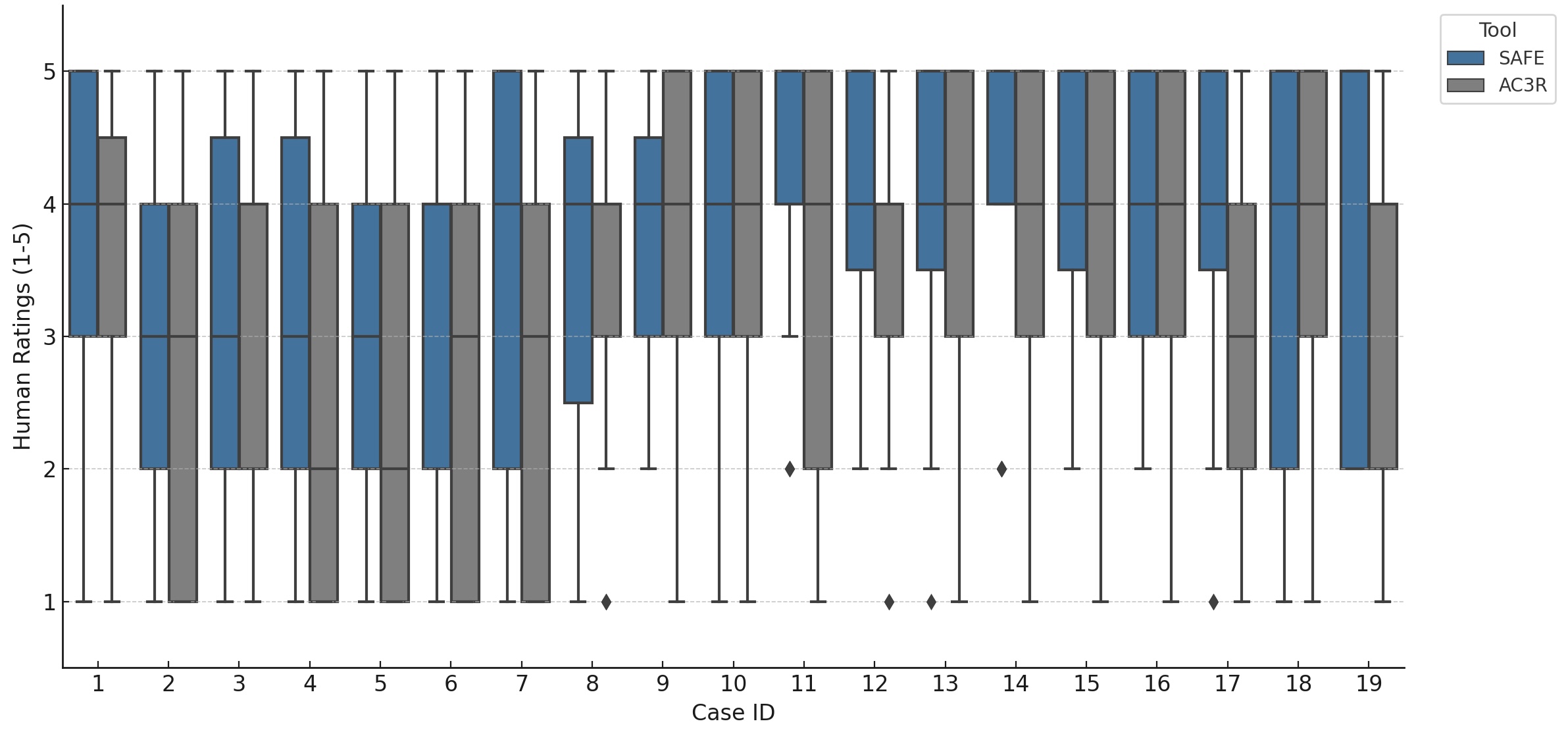}
      \Description{This subfigure presents the human evaluation of road network reconstruction consistency between SAFE and AC3R, comparing how accurately each method reconstructs road structure.}
      \caption{Road Network Alignment Comparison}
      \label{fig:subfig_b_1}
  \end{subfigure}
  
  \begin{subfigure}{\linewidth}
      \centering
      \includegraphics[width=0.9\linewidth]{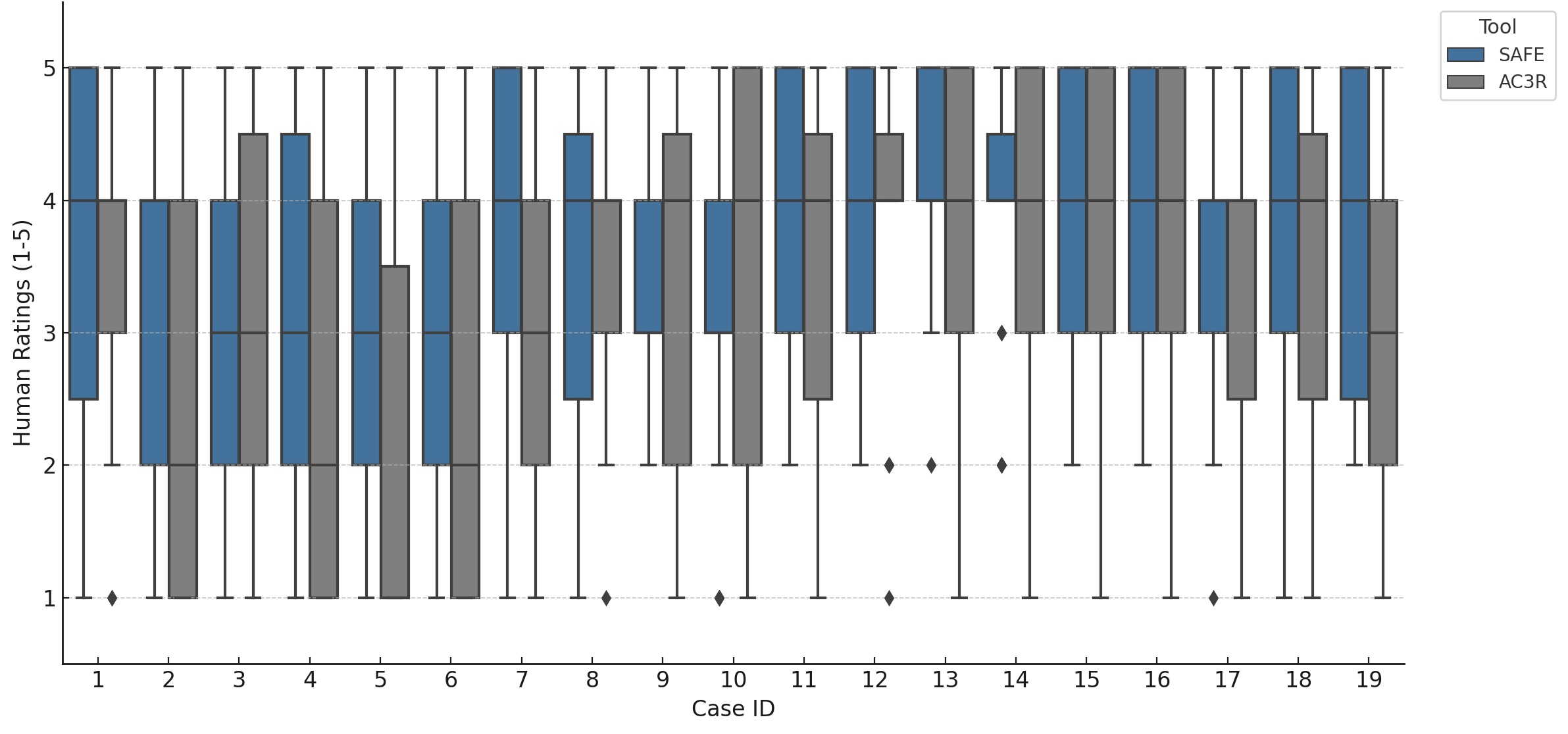}
      \Description{This subfigure presents the human evaluation of actor behaviour reconstruction consistency between SAFE and AC3R, assessing how well each method reproduces vehicle movements and interaction patterns.}
      \caption{Actor Behaviours Alignment Comparison}
      \label{fig:subfig_b_2}
  \end{subfigure}
  \caption{{\tool} vs. {\toolthree} Human Study Results}
  \label{fig:safe_ac3r_human_study}
\end{figure}

Figure~\ref{fig:safe_ac3r_human_study} shows the questionnaire results comparing \tool{} and \toolthree{}. To assess overall performance, we conducted Mann–Whitney U tests on ratings from 35 participants across 19 cases. For Road Network Alignment, $U=245{,}772.5$ with $p=0.00027$, and for Actor Behavior Alignment, $U=242{,}945.0$ with $p=0.00133$, both indicating statistically significant differences in favor of \tool{}.

\begin{figure}[!ht]
  \centering
  \includegraphics[width=\linewidth]{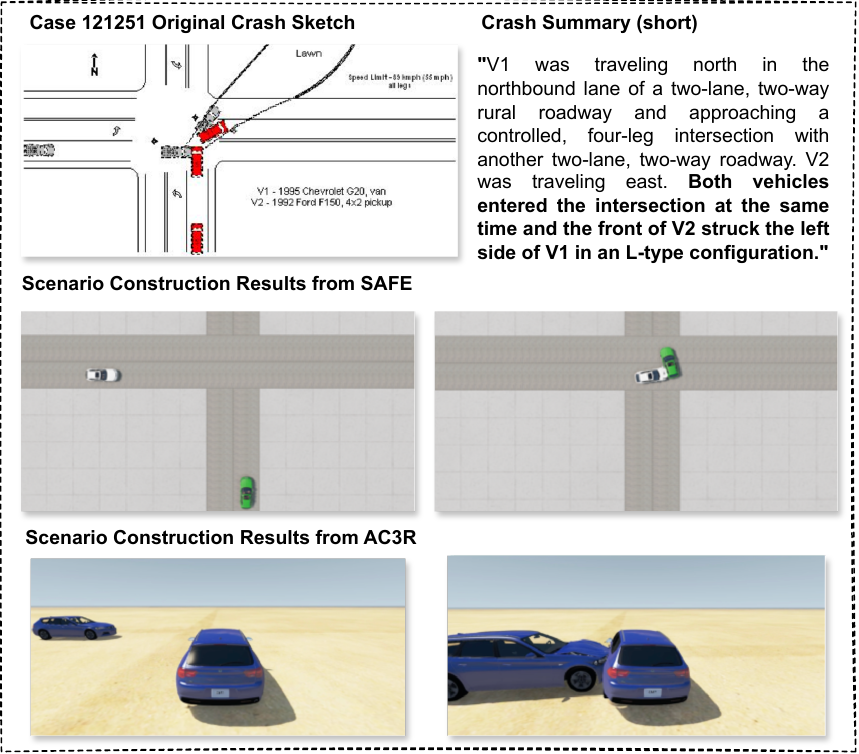}
  \Description{This figure shows SAFE V.S. AC3R on the crash case 121251}
  \caption{Scenario Construction from SAFE and AC3R for Case 121251}
  \label{fig:sc_com_3}
\end{figure}

In the comparison between \tool{} and \toolthree{}, certain cases—such as Cases 3, 9, 10, and 14—show \toolthree{} outperforming \tool{} in average actor alignment scores. For example, in Case 14 (Crash Report 121251), as shown in Figure~\ref{fig:sc_com_3} row 1, both vehicles entered the intersection simultaneously, with V2 striking the left side of V1 in an L-shaped collision.
In the simulation results from \tool{} (Figure~\ref{fig:sc_com_3}, row 2), V2 strikes the rear-left of V1, whereas \toolthree{} (row 3) shows V2 impacting the mid-front left side—closer to the original report’s description of ``the front of V2 struck the left side of V1 in an L-type configuration.'' While both tools received an average score around 4 (``Mostly Match''), \toolthree{} had more perfect scores of 5 (``Totally Match''). Analysis of \tool{}’s outputs revealed the error stemmed from failing to capture fine-grained initial vehicle positions, particularly their relative distances from the intersection. This case highlights a limitation of \tool{} in preserving precise spatial details, which can impact reconstruction accuracy. Nonetheless, as shown in Figure~\ref{fig:safe_ac3r_human_study}, \tool{} still achieves higher overall performance based on average user ratings.

% https://drive.google.com/file/d/1nx-_yp9hG68cKQHKhwjukwpw-tkgdidr/view?usp=sharing

\subsection{RQ3: Effectiveness in Scenario Construction, ADS Bug Detection and Crash Reproduction}
We evaluated {\tool}'s scenario generation and bug detection across MetaDrive and BeamNG simulation platforms with various ADS systems (IDM, PPO, Auto). Table~\ref{tab:ADStesting} summarizes the scenario construction and ADS bug detection performance of \tool{} and the baselines across 50 crash reports, with results averaged over five runs. Compared to \tooltwo{} under the same simulator (MetaDrive) and ADS model (IDM), \tool{} generates 164 more scenarios and detects 39 more violations. A U-test yields a p-value of 0.0109, confirming at the 95\% confidence level that \tool{} is significantly more effective in exposing ADS bugs.

Against \toolthree{}, which only supports 19 of the 50 reports due to its reliance on traditional NLP and fixed ontologies, \tool{} generates 77 more scenarios and detects 71 additional violations (p-value = 0.0099). These results also indicate a statistically significant improvement.

Table~\ref{tab:bug_counts} reports the number of scenarios required to trigger Top-K violations, further demonstrating \tool{}’s efficiency in generating targeted test cases. \tool{} requires an average of 10 seconds per scenario in MetaDrive and 25 seconds in BeamNG, with an overall violation detection rate of 30.52\%. By comparison, \tooltwo{} averages 4 seconds per case (due to lack of simulator rendering) but has a lower detection rate of 6\%. \toolthree{} averages 27 seconds per case and achieves a high detection rate of 84.21\%, but lacks generalizability and is incompatible with MetaDrive.

\begin{table}[h]
    \centering
    \label{tab:ads_results}
    \resizebox{\columnwidth}{!}{
    \begin{tabular}{p{2.8cm}cccccc}
        \hline
        \multirow{2}{*}{\textbf{ADS \& Simulator}} & \multicolumn{3}{c}{\textbf{\# Scenarios}} & \multicolumn{3}{c}{\textbf{\# Violations}} \\
        \cline{2-7}
        & SAFE & LCTGen & AC3R & SAFE & LCTGen & AC3R \\
        \hline
        PPO MetaDrive & \textbf{211} & N/A & N/A & \textbf{30} & N/A & N/A \\
        IDM MetaDrive & \textbf{214} & 50 & N/A & \textbf{42} & 3 & N/A \\
        Auto BeamNG & \textbf{96} & N/A & 19 & \textbf{87} & N/A & 16 \\
        \hline
    \end{tabular}}
    \caption{Performance Comparison in Scenario Construction \& Testing}
    \label{tab:ADStesting}
\end{table}

\begin{table}[!t]
\centering
\scalebox{0.8}{ 
\begin{tabular}{llr}
\toprule
\textbf{Tool - Simulator - ADS} & \textbf{Level} & \textbf{Count} \\
\midrule
\multirow{3}{*}{{\tool} - MetaDrive - IDM} & Top 1 - violation & 1 \\
                                         & Top 2 - violation & 3 \\
                                         & Top 3 - violation & 4 \\
\midrule
\multirow{3}{*}{{\tool} - MetaDrive - PPO} & Top 1 - violation & 3 \\
                                         & Top 2 - violation & 11 \\
                                         & Top 3 - violation & 12 \\
\midrule
\multirow{3}{*}{{\tool} - BeamNG - Auto}          & Top 1 - violation & 1 \\
                                         & Top 2 - violation & 2 \\
                                         & Top 3 - violation & 3 \\
\midrule
\multirow{3}{*}{{\tooltwo} - MetaDrive - IDM}          & Top 1 - violation & 6 \\
                                         & Top 2 - violation & 17 \\
                                         & Top 3 - violation & 29 \\
\midrule
\multirow{3}{*}{{\toolthree} - BeamNG - Auto}          & Top 1 - violation & 2 \\
                                         & Top 2 - violation & 5 \\
                                         & Top 3 - violation & 7 \\            
\bottomrule
\end{tabular}}
\caption{Violations Counts by Level for Different Tools}
\label{tab:bug_counts}
% \vspace{-6mm}
\end{table}
% -----------------------------------------
% \begin{figure}[t]
%   \centering
%   \begin{subfigure}{\linewidth}
%       \centering
%       \includegraphics[width=0.9\linewidth]{Graphs/case_1.jpg}
%       \caption{Violation on T-intersection road}
%       \label{fig:3_subfig1}
%   \end{subfigure}
  
%   % \vspace{1em}  % Add vertical space between subfigures
  
%   \begin{subfigure}{\linewidth}
%       \centering
%       \includegraphics[width=0.9\linewidth]{Graphs/case_2.jpg}
%       \caption{Violation on merging road}
%       \label{fig:3_subfig2}
%   \end{subfigure}
  
%   % \vspace{1em}  % Add vertical space between subfigures
  
%   \begin{subfigure}{\linewidth}
%       \centering
%       \includegraphics[width=0.9\linewidth]{Graphs/case_3.jpg}
%       \caption{Violation on intersection road}
%       \label{fig:3_subfig3}
%   \end{subfigure}
  
%   \caption{Three Founded Violations in MetaDrive}
%   \label{fig:three_cases}
% \end{figure}
% -----------------------------------------
% Modify to horizontal arrangement
\begin{figure}[t]
  \centering
  \begin{subfigure}[b]{0.31\linewidth}
      \centering
      \includegraphics[width=\linewidth]{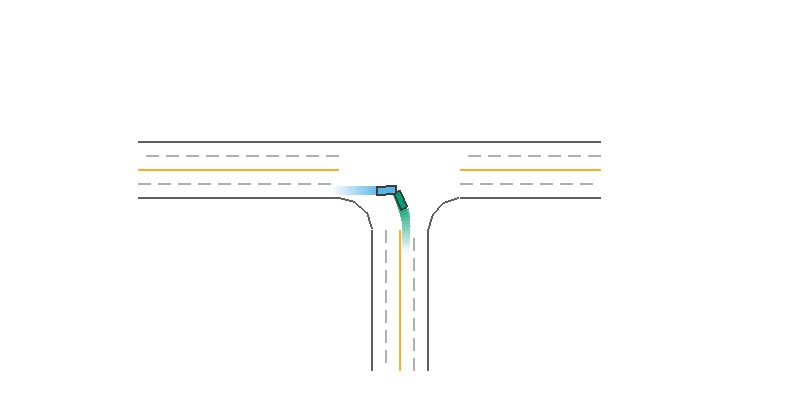}
      \Description{A violation found on a T-intersection road}
      \caption{T-intersection}
      \label{fig:3_subfig1}
  \end{subfigure}
  \hfill
  \begin{subfigure}[b]{0.31\linewidth}
      \centering
      \includegraphics[width=\linewidth]{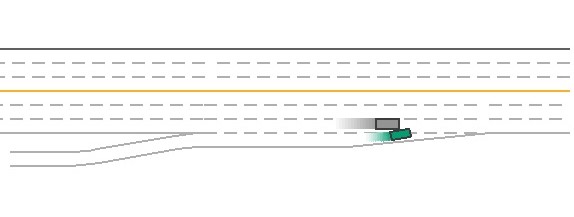}
      \Description{A violation found on a merging road}
      \caption{Merging road}
      \label{fig:3_subfig2}
  \end{subfigure}
  \hfill
  \begin{subfigure}[b]{0.31\linewidth}
      \centering
      \includegraphics[width=\linewidth]{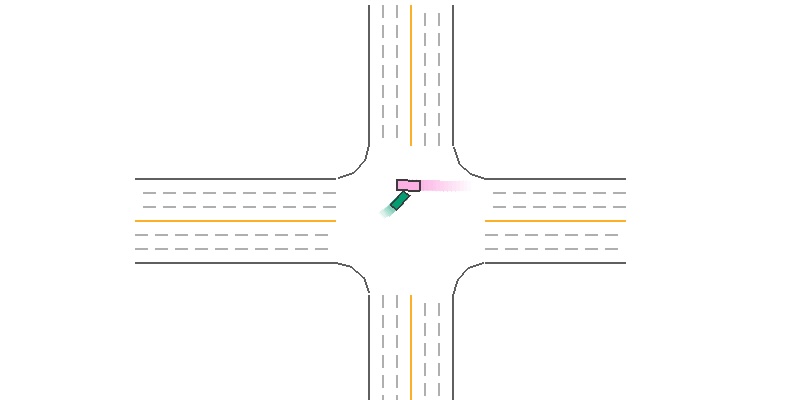}
      \Description{A violation found on an Intersection road}
      \caption{Intersection}
      \label{fig:3_subfig3}
  \end{subfigure}
  \caption{Three founded violations in MetaDrive}
  \label{fig:three_cases}
\end{figure}
% -----------------------------------------
Figure ~\ref{fig:three_cases} showcases three representative violations detected by {\tool} in the MetaDrive simulator. In Figure ~\ref{fig:3_subfig1}, the ADS vehicle fails to properly yield at a T-intersection, colliding with a vehicle that has the right of way. Figure ~\ref{fig:3_subfig2} illustrates a violation where the ego vehicle attempts to merge onto a highway from an on-ramp but fails to yield to the vehicle already on the highway, resulting in a side collision. Figure ~\ref{fig:3_subfig3} shows a violation at an intersection where the ego vehicle making a turn fails to yield to an opposing vehicle traveling straight, leading to a collision. These diverse violation types demonstrate {\tool}'s effectiveness in generating realistic and challenging test scenarios based on traffic accident reports, which exposes various weaknesses in ADS decision-making and control algorithms across different road configurations and traffic conditions.

\begin{figure}[t]
  \centering
  \begin{subfigure}{\linewidth}
      \centering
      \includegraphics[width=0.9\linewidth]{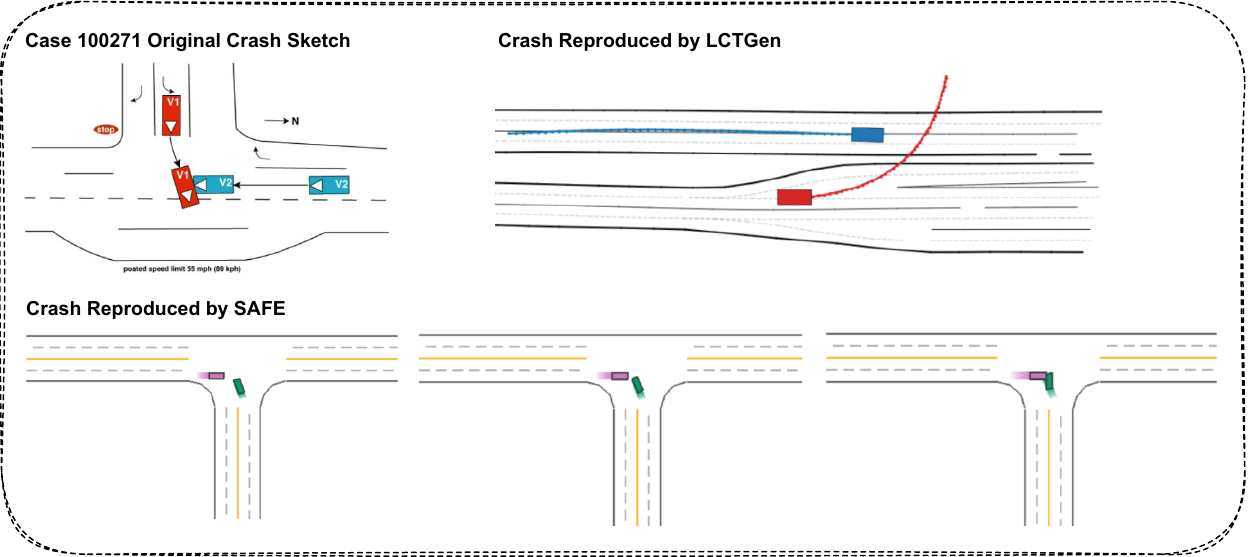}
      \Description{Reconstruction on the case 100271}
      \caption{Crash Report - 100271}
      \label{fig:subfig1}
  \end{subfigure}
    
  \begin{subfigure}{\linewidth}
      \centering
      \includegraphics[width=0.9\linewidth]{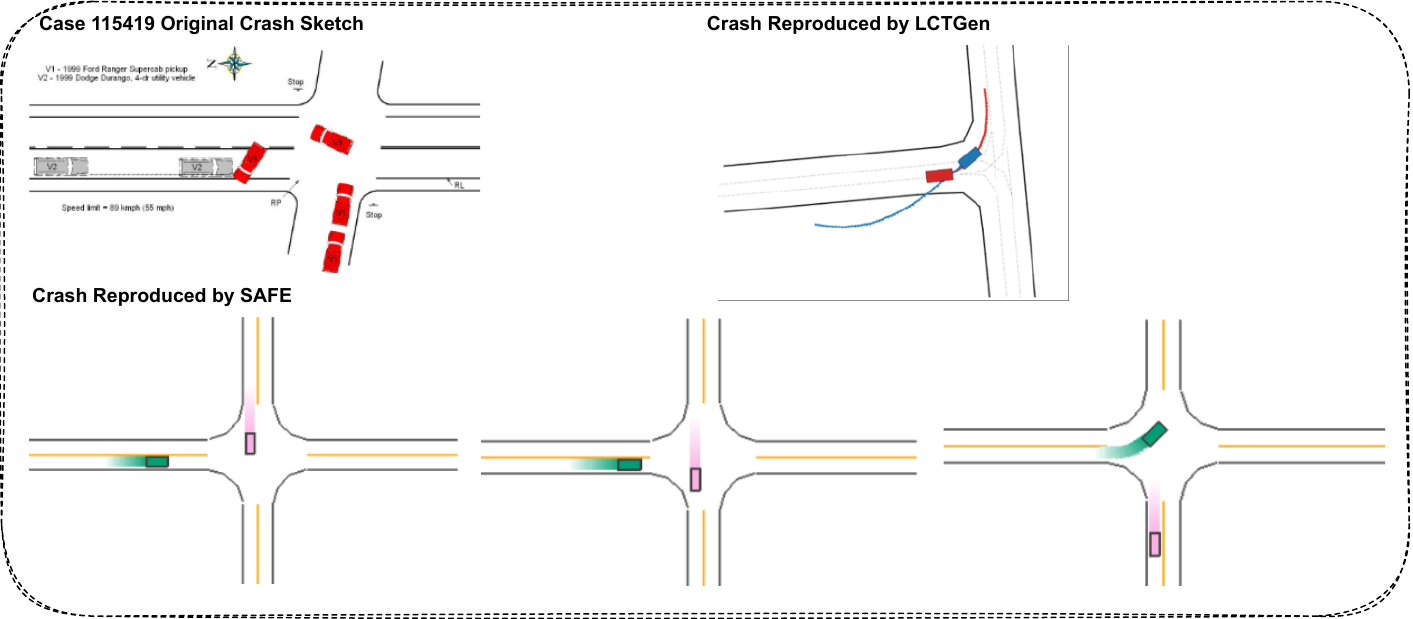}
      \Description{Reconstruction on the case 115419}
      \caption{Crash Report - 115419}
      \label{fig:subfig2}
  \end{subfigure}
  \caption{{\tool} vs. {\tooltwo} on Crash Reproduction}
  \label{fig:safe_lctgen_crash_reproduce}
\end{figure}

Regarding the evaluation of crash reproduction ability between {\tool} and {\tooltwo}, five independent runs show that {\tool} successfully reproduced 15 crashes on average, whereas {\tooltwo} only managed to reproduce 3. The U-test results on the number of reproduced crashes yield a p-value of 0.0099, suggesting that {\tool} is significantly more effective at generating crash-inducing scenarios compared to {\tooltwo} under this experimental setting. Figures ~\ref{fig:subfig1}, and ~\ref{fig:subfig2} show the reproduction results of {\tool} and {\tooltwo} based on two different crash reports (100271 \& 115419). 

In Figure ~\ref{fig:subfig1}, {\tool} successfully reconstructed both the road network and vehicle behaviors. In contrast, {\tooltwo} only reconstructed the correct vehicle trajectories but failed to accurately restore the road structure, which led to an incorrect reconstruction of the crash. In Figure ~\ref{fig:subfig2}, we can see that although {\tool} did not reproduce the crash described in the report, it successfully reconstructed the road structure. As shown in Figure ~\ref{fig:subfig2} {\tool}’s simulation result, although both vehicles performed the actions described in the crash report—one driving straight through the intersection and the other making a turn—the two vehicles did not encounter each other at the same time, and thus the crash was not reproduced. For {\tooltwo}'s result, while it successfully constructed a crash case, it reconstructed the road network incorrectly.

In evaluating crash reproduction on 19 reports, \tool{} successfully reproduced 17 crashes on average across five runs, while \toolthree{} reproduced 16. A U-test yields a p-value of 0.0189, indicating a statistically significant difference (p < 0.05) in favor of \tool{}. Moreover, \tool{} generalizes to a broader range of data and generates more realistic test cases, as it does not rely on a fixed ontology and performs well on unseen scenarios.

\begin{figure}[t]
  \centering
  \includegraphics[width=\linewidth]{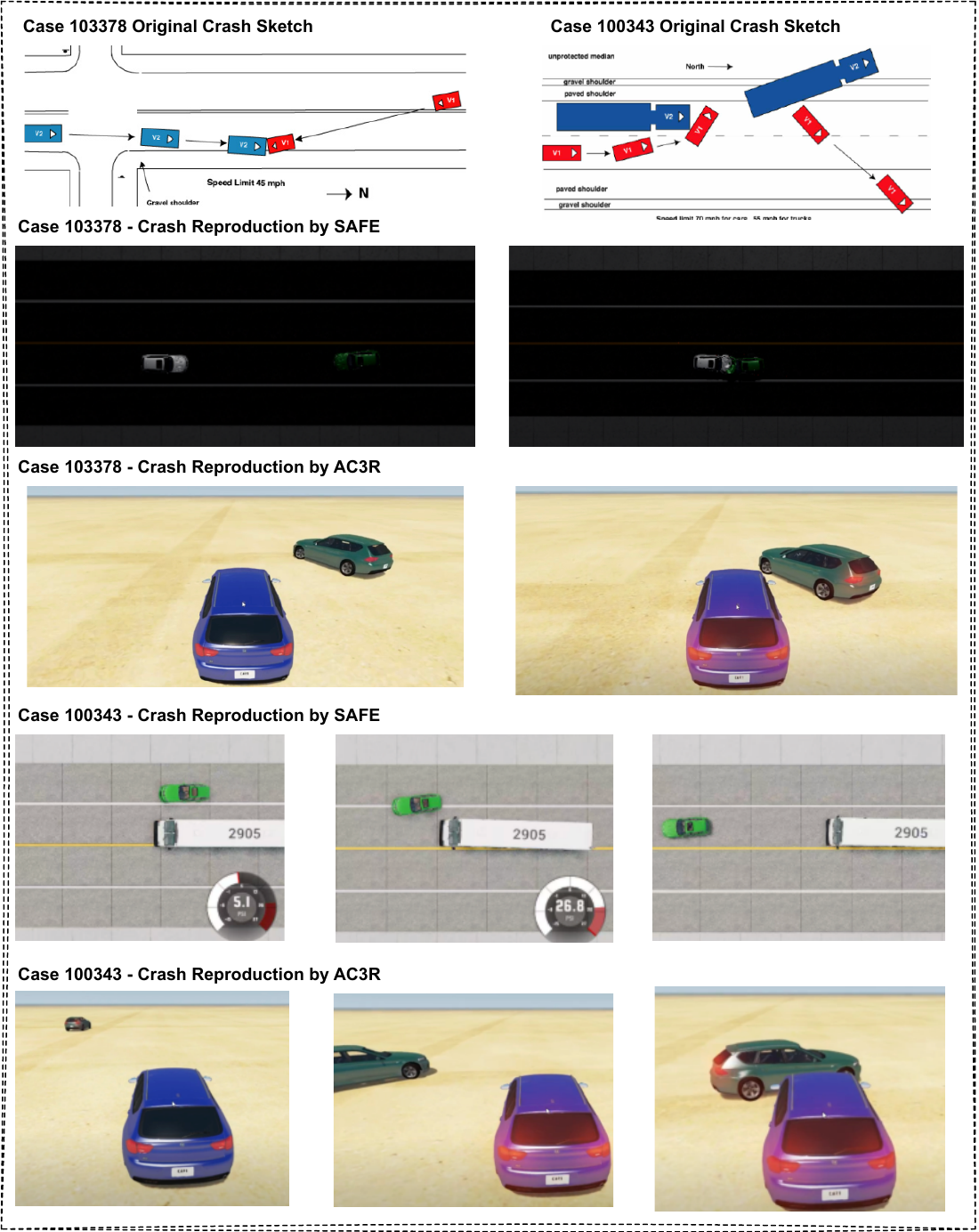}
  \Description{the comparison between SAFE and AC3R on Crash Reproduction }
  \caption{{\tool} vs. {\toolthree} on Crash Reproduction}
  \label{fig:safe_ac3r_crash_reproduce}
\end{figure}

Figure ~\ref{fig:safe_ac3r_crash_reproduce} shows the reproduction results of {\tool} and {\toolthree} on two different crash reports (103378 \& 100343). From the Crash Reproduction by {\tool} of Case 103378 as shown in Figure ~\ref{fig:safe_ac3r_crash_reproduce}, we can see that the road network is successfully reconstructed, vehicle actions are accurately reproduced, the crash is detected, and even the dark weather condition is reflected. Clear lane markings are also visualized. In contrast, for {\toolthree}’s result, it does not support front-end crash cases and instead reconstructs them as sideswipe crashes. While {\toolthree} successfully generates an intersection in the road network, the road markings are unclear, and the lanes appear too narrow. For the Crash Report 100343 reproduction, as shown in the figure ~\ref{fig:safe_ac3r_crash_reproduce}, {\tool} successfully reconstructs the road network, vehicle types, and behaviors. From the Crash Reproduction by {\toolthree} of Case 100343, it shows {\toolthree} did not effectively capture the information from the crash reports.

% https://drive.google.com/file/d/1nx-_yp9hG68cKQHKhwjukwpw-tkgdidr/view?usp=sharing

% Combine Figure 11 & 12 SAFE V.S. AC3R

\subsection{RQ4: Effectiveness of Prompt Generation and Self-Validation}
To investigate the effectiveness of {\tool}'s prompt generation and self-validation components, we conducted ablation studies by disabling these two methods and observing the resulting impact on scenario representation extraction accuracy. 

\begin{table}[]
\begin{tabular}{lcc}
\hline
\textbf{}    & \textbf{\begin{tabular}[c]{@{}c@{}}prompt generation \\ removal\end{tabular}} & \textbf{\begin{tabular}[c]{@{}c@{}}prompt generation and \\ self-validation removal\end{tabular}} \\ \hline
Road Network & 77.0\%                                                                          & 77.0\%                                                                                              \\
Actors       & 18.0\%                                                                        & 18.0\%                                                                                            \\
Env          & 87.1\%                                                                          & 84.73\%                                                                                              \\
Overall      & 10.3\%                                                                         & 12.8\%                                                                                             \\ \hline
\end{tabular}
\caption{Performance of {\tool}'s Stage I after Removing Prompt Generator and Self-validator}
\label{tab:ablation}
% \vspace{-6mm}
\end{table}

Table ~\ref{tab:ablation} shows the average results of three independent replicate experiments. When the Prompt Generation mechanism is disabled, {\tool}'s mean accuracy on the road network extraction decreased significantly from 93.8\% to 77\%, accuracy on actors decreased from 80.0\% to 18.0\%, and environment accuracy decreased from 100\% to 87.1\%, as shown in Table~\ref{tab:ablation}. Table ~\ref{tab:rm_pg_re} presents the information extraction results of {\tool} for Case 119489 (the sketch is shown in Figure ~\ref{fig:sc_safe_lctgen}) without the prompt generator. The results indicate that {\tool} produced incorrect vehicle location information in actor extraction. For road network extraction, it did not accurately identify the road type but merely referenced the road characteristics described in the crash summary. When we further removed the self-validator, the environment accuracy decreased further to 84.73\%, and the overall accuracy changed to 12.8\%. These results highlight the critical role of both components in {\tool}'s performance. The dramatic drop in actor extraction accuracy suggests that the Prompt Generation mechanism is particularly essential for identifying and correctly characterizing dynamic elements in the scenario. The more modest decreases in road network and environment accuracy indicate that these static elements are somewhat easier to extract, though still benefit significantly from structured prompting. Self-verification appears to provide an additional layer of accuracy for environmental elements, likely by catching and correcting initial extraction errors through its validation process.

\begin{table}[]
\centering
\resizebox{\columnwidth}{!}{
\begin{tabular}{p{0.9\columnwidth}}
\toprule
\rowcolor{gray!20} SAFE's Respond \\
\#\#\# Explanation:

- \textbf{Actors}:

  - \textbf{Vehicle 1 (V1)}: 
    - Model: Sedan
    - Initial position: W2E (waiting to turn left)
    - Actions: Attempting to turn left
    - Speed: Not specified

  - \textbf{Vehicle 2 (V2)}: 
    - Model: Sedan
    - Initial position: N2S (moving north in the northbound lane)
    - Actions: Moving forward
    - Speed: 45 mph (based on posted speed limit)

- \textbf{Road Network}:
  - Described as a north-south two-lane, two-way road.

- \textbf{Env}:
  - Time: Daytime
  - Weather: Snowy, with icy road conditions.\\
\bottomrule
\end{tabular}%
}
\caption{Results from {\tool} without Prompt Generator on Case 119489}
\label{tab:rm_pg_re}
\vspace{-8mm}
\end{table}

%% file: Discussion.tex
\section{Validity Discussion and Future Work}
\label{sec:discuss}
\noindent\textbf{Validity} We tested across two off-the-shelf simulators and three representative end-to-end ADS architectures but did not evaluate on CARLA or the multi-module ADS systems Baidu Apollo~\cite{apollo} and Autoware~\cite{autoware}. This is because our chosen simulators rely on built-in map construction capabilities, whereas CARLA primarily offers predefined maps, making reconstruction difficult. Additionally, both Apollo and Autoware are supported by CARLA, but not supported by our selected simulators. To mitigate this limitation, we tested three different end-to-end ADS systems compatible with our simulators. We did not test other SOTA LLM-based scenario generation methods listed in the related work, as they are no longer supported and cannot be reproduced. However, we reproduced another SOTA LLM-based approach, LCTGen, along with AC3R, the original work that inspired this direction.

\noindent\textbf{ Extensibility}
SAFE is a flexible, modular framework designed to generalize across diverse datasets and construct complex traffic scenarios. It operates in two stages. In the Information Extraction stage, plug-and-play modules—the Meta Message Extractor, Prompt Generator, Self-Validator, and Information Encoder—convert a multimodal crash report into a structured scenario expressed in a DSL. Our DSL extends TARGET by retaining its three core components (Road Network, Traffic Actors, and Weather) while allowing users to add custom fields for scenario-specific details. The DSL also supports replacing high-level actor behaviours with precise coordinate sequences, enabling RAG-constrained LLMs to generate more accurate trajectories and improving crash reproduction efficiency.

In the Scenario Construction and Testing stage, a simulator-specific Adapter compiles the DSL into executable test cases. SAFE generates diverse, semantically consistent scenarios by sampling valid placements in MetaDrive and physically plausible parameters in BeamNG, allowing multiple meaningful test cases to be derived from a single crash report. Our evaluation expands the LCTGen dataset with 50 NHTSA crash reports and demonstrates successful simulation in both MetaDrive and BeamNG. Although we implement adapters for these two simulators, SAFE is designed to be readily extensible to others.

\noindent\textbf{Future Work} 
%Based on SAFE’s performance in scenario representation extraction, scenario construction and ADS bug detection, and crash reproduction, we have identified two main limitations. 
From SAFE’s performance, we identified two key limitations. First, SAFE occasionally fails to accurately interpret visual details in multimodal crash reports, which may lead the Prompt Generator to select incorrect templates and produce faulty scenario representations. This can result in either failed scenario reconstruction or successful reconstruction that does not reproduce the critical elements described in the crash report. Second, although we have applied techniques such as few-shot learning, CoT prompting, and self-validation to mitigate hallucinations, LLMs may still exhibit occasional hallucinations due to their inherent non-determinism and stochastic nature. To address these limitations, we believe that a more fine-grained DSL, along with annotated datasets, is needed to support the use of RAG, which can help constrain the LLM’s output based on factual evidence.

% Our evaluation employs four core accuracy metrics—Road Network, Actor, Environment, and Overall Accuracy—which directly align with the modular architecture of SAFE and its prompt design.
Our evaluation employs four core accuracy metrics, which directly align with the modular architecture of SAFE and its prompt design. These metrics provide a balanced trade-off between interpretability and comprehensiveness, capturing the essential elements for scenario reconstruction. Nonetheless, we acknowledge that our current metrics do not explicitly account for hallucinated or extraneous objects, which may impact the perceived fidelity of reconstructed scenarios. Addressing this limitation remains an important direction for future work.

%% file: Conclusion.tex
\section{Conclusion}
\label{sec:conclusion}
We presented \tool, a novel framework that leverages LLMs to generate ADS test scenarios from multimodal crash data. To mitigate hallucination, we introduced a compositional DSL with verifiable constructs and a RAG-based knowledge base that provides component-level prompting. We further proposed two map reconstruction pipelines—map synthesis in BeamNG and map adaptation in MetaDrive—enabling flexible deployment across simulators.

\tool\ outperforms existing approaches in knowledge extraction accuracy, scenario fidelity, and ADS safety issue detection. As one of the first systems to couple multimodal crash data with LLM-driven scenario generation, it lays the groundwork for future research on multimodal corner-case synthesis~\cite{10.1145/3663529.3663779} and neurosymbolic specification extraction~\cite{zheng2025neurostrataharnessingneurosymbolicparadigms}.
%In this paper, we introduce {\tool}, one of the first approaches to generate ADS test scenarios from multimodal crash data using LLMs. To mitigate hallucination, we design a compositional and easily validated DSL and incorporate a knowledge base following the RAG framework, pre-storing prompts for each DSL component to provide a chain of thought.

%For each DSL representation, we employ two map reconstruction methods based on the chosen simulators: building maps from scratch in BeamNG and adapting existing maps in MetaDrive. Our results demonstrate significant improvements over SOTA methods in LLM knowledge extraction accuracy, crash scenario fidelity, and ADS safety issue detection.

%This work is significant not only as the first to integrate LLMs with multimodal data for ADS testing but also as a foundation for generating scenarios encompassing both dynamic object trajectories (e.g., vehicles) and road networks. It establishes a solid basis for promising research on LLM-driven corner case generation from multimodal data~\cite{10.1145/3663529.3663779} and the extraction of formal specifications using the neurosymbolic paradigms~\cite{zheng2025neurostrataharnessingneurosymbolicparadigms}.